\documentstyle[11pt,epsf,paspconf]{article}

\begin{document}

\title{Constraining Dark Halo Potentials with Tidal Tails}

\author{John Dubinski}
\affil{CITA, University of Toronto, 60 St. George St., Toronto,~ON~M5S~3H8,
Canada}
\author{Lars Hernquist}
\affil{Board of Studies in Astronomy and Astrophysics, 
UC Santa Cruz, Santa~Cruz,~CA~95064}
\author{J. C. Mihos}
\affil{ Department of Astronomy,
        Case Western Reserve University,
        10900~Euclid~Ave,
        Cleveland,~OH~44106}





\begin{abstract}
Massive and extended dark halos can inhibit the formation of long tidal tails 
in galaxy collisions.  We examine this effect 
using an extensive survey of simulations with different dark halo potentials
to constrain halo properties of interacting galaxies.  These constraints 
are compared to other observational limits and theoretical predictions 
of halo structure.  The dark halos predicted by $\Omega=1$ cosmological
models like CDM are too massive and extended to produce the long tidal
tails seen in nearby galaxy collisions.  There is also a conflict with the halo
potentials inferred from satellite kinematics; such halos would likewise
inhibit tail formation in galaxy collisions.
\end{abstract}


\keywords{galaxy dynamics: interacting galaxies: dark matter: cosmology}


\section{Introduction}

When disk galaxies collide, the strong mutual tidal fields of their interaction
momentarily transform the disks into open, bisymmetric spirals and catapult 
the outer disk stars in each galaxy 
onto long, arcing trajectories commonly called tidal
tails and bridges (Toomre \& Toomre 1972; Wright 1972).  
Tidal bridges connect the merging pair and
tidal tails continue to lengthen and thin out after a collision.
Some tail stars may escape the system but most
eventually fall back into the merger remnant.
The currently
interacting pairs in the Antennae (NGC 4038), the Mice (NGC 4676) and the
merged NGC 7252 for example, all have tails which extend to distances
$\sim$50-100 kpc from the galaxies (Hibbard 1995).
The Superantennae 
(IRAS 19254-7245) is an extreme case in which the tails span  
350 kpc from tip to tip
(e.g., Mirabel, Lutz, \& Maza 1991).

A tidal tail can be thought of as a single trajectory shaped by the 
potential of the interacting pair since it arises from stars in the disk
with similar orbital properties.
Tidal tails can extend far out into the dark halo,
perhaps out to $>$200 kpc in 3 dimensions.  
Tails were therefore recognized as a possible
probe of the dark matter distribution in galaxies (Faber and Gallagher
1979) and their formation in simulations with dark halos has been the subject
of much numerical work during the past decade (e.g. Negroponte \& White
1983; White 1982; Barnes 1988). 
The suggestion is that massive dark halos can inhibit the formation of tidal
tails
in two ways.  First, the deeper potentials could lead to larger relative
encounter velocities which are more impulsive and less resonant and therefore
less effective at giving disk stars the velocity perturbation they need
to be ejected as tails.  Second, the deeper and steeper potential wells of
more massive halos could trap tail stars by shortening their
turn-around radius after ejection.
Both effects would combine to make the maximum length of tidal tails
smaller in the presence of a more massive dark halo.  The observed {\em maximum}
length of tails is therefore telling us something about the mass profile 
of the dark potential.


The length and kinematics of tidal tails
depend on other factors besides the dark halo structure
so simulation surveys are required to understand how they form in different
circumstances.
Other factors
include the epoch of the collision, disk orientations, orbit orientation, 
orbital energy, galaxy mass ratio, and pericentric distance.
The only way to study dark halo potential constraints is therefore
to model the tail
formation process as a function of dark halo parameters in galaxy
collision N-body simulations (Dubinski, Mihos \& Hernquist 1996 [DMH96]).
A complete parameter survey is out of reach so in
our studies we have focussed on the ideal case of colliding  
equal mass, co-planar disks in a direct encounter.   This is the geometry
of the first galaxy collision ever simulated, namely that done
by Holmberg in 1941 ({\em
before} the computer age!). 
These encounters are strongly resonant and
are the most effective at transferring energy and angular momentum to the tails
and so the lengths of tails in these collisions 
probably represent an upper limit for all collision geometries.
We have also examined more specific geometries which would produce
facsimiles of the Antennae and NGC 7252 using different mass models
(DMH96; Mihos, Dubinski, Hernquist 1998).

\section{Previous Results}

In previous work, we simulated galaxy collisions
using four mass models labelled A, B, C and D
with progressively more massive and more extended halos (DMH96).
The models are self-consistent realizations containing
an exponential disk, a truncated King model bulge and King model dark halo, 
constructed using the method of Kuijken \&
Dubinski (1995).  
By construction, the disk and bulge mass distributions are held fixed and
the inner rotation curve is kept nearly flat within 5 scale lengths in the 
model sequence.
With this assumption, the mass ratio of dark matter to stars (disk + bulge)
is 4:1, 8:1, 16:1 and 30:1 for models A, B, C, and D.
Of course, more extended models with this mass ratio are also possible
when dark halos have smaller circular velocities
but we have only
examined those with rotation curves which are approximately flat out to the
halo truncation radius.

The results of this study are given in DMH96 and the main points are
summarized here.
Galaxy collisions with models A and B have no difficulty 
ejecting long tails  ($l > 100$ kpc)
while model C 
encounters eject tails of only moderate length ($l \sim 50-100$ kpc)
and model D collisions produce short tails ($l < 50$ kpc).
When these simulations are carried through to the final merger to
a state similar to NGC 7252, models A and B still have long tails
while all of the stars in the tails in model C and D
galaxies have fallen back into the merger remnant.  
On these grounds, we can rule out galaxies with
mass distributions like models C and D as precursors of the Antennae, the
Mice and NGC 7252.
Collision geometries that produce facsimiles of the Antennae and NGC 7252
worked well with the low mass models A and B but failed with models C and D.
Also, kinematic comparisons of NGC 7252 facsimiles to the real
observations
also rule out model C and D even when we include an ``HI gas'' disk
(represented by test particles) extending out to 10 scale lengths (MDH98).
Since models C and D had dark halo mass
distributions very similar to those predicted from CDM or other
$\Omega=1$ cosmological models, we argued that a
lower $\Omega$ cosmology might be required to explain the observations.  
Indeed, small $\Omega$
cosmologies produce concentrated dark halos
that look more like those in models A and B (Navarro et al. 1997).


\section{New Results}

How representative are these 4 models?  They were constructed under the
assumption that the flat inner rotation curve continues to large
radii.  Clearly, there are many other possibilities.  For example, at large
radii in an
isothermal halo, $\phi \sim v_c^2 \log r$ so $\partial \phi/\partial r \sim
v_c^2/r$.  If the disk
contributes considerably more mass to the inner rotation curve the {\em
asymptotic} value of $v_c$ may be smaller and so the outer potential will
be shallower.  Tidal tails may therefore be expelled
to larger distances in this case
since the potentials are not as effective a trap.  
This is just one example of how things can
stray from our first 4 models.

The only way to understand the effects on tidal tail formation is 
to widen the survey to include other dark halo potentials.  
We have done just that by looking
at 84 new models (Dubinski, Mihos, Hernquist [DMH98]).
We have
parameterized the dark halo using two models which have been motivated by
cosmological simulations of dark halo formation: the Hernquist (H) profile
(Hernquist 1990; Dubinski \& Carlberg 1991) and the NFW profile (Navarro, 
Frenk \& White 1996).  Both profiles have two free parameters, a mass (or
effective mass within some radius 
in the case of the NFW profile since total mass formally diverges)
and a scale length.  The H-profile is:
\begin{equation}
\rho(r) = \frac{M_H r_H}{2 \pi} \frac{1}{r(r+r_H)^3},
\end{equation}
where $M_H$ is the H mass and $r_H$ is the H scale length.
The NFW profile is:
\begin{equation}
\rho(r) = \frac{M_s}{4 \pi} \frac{1}{r(r+r_s)^2},
\end{equation}
where $M_s$ is the NFW effective mass (mass within 5.3 $r_s$ )
and $r_s$ is the NFW scale length.
Both profiles have peak circular velocities which we use to characterize
them.  For the H model $v_H^2 = 0.5 GM_H/r_H$ at $r=r_H$ and for the NFW
model $v_s^2 = 0.46 G M_s/r_s$ at $r=2.16 r_s$.  The velocity maxima of the
two dark halo rotation curves coincide when we use 
the mapping $r_s = 0.46 r_H$ and $M_s = 0.54 M_H$.
The NFW model is more extended than the H model and has a steeper
potential at larger radii, however, within the region where tails form
the potentials are very similar.  We found that there was very little
difference in the results for both models so we only present the results of
the NFW models here, although a more extensive comparison will be given
elsewhere (DMH98).

In our previous work, we simulated the collisions with self-consistent
treecode N-body simulations.  The large number of models in this
extended study precludes the use of self-consistent simulations  at the
moment
so we use instead a 
restricted, 3-body method somewhat similar to Toomre \& Toomre's (1972)
original scheme.  We can get away with this since the tidal tails
are essentially a kinematic phenomenon.  The technique will be described
elsewhere in detail (DMH98) but in brief it works
as follows.  The trajectories of the two interacting galaxies are
calculated assuming an interaction potential with zero total energy
which accounts for the
extended mass distribution and is somewhat different than the parabolic
orbits of point masses used by Toomre \& Toomre.  
The orbital decay of the galaxies is treated used
Chandrasekhar's dynamical friction formula (Binney \& Tremaine 1987) 
using a value of the Coulomb
logarithm, $\ln \Lambda = 2$, found to fit the observed orbital decay well in
a small number of 
self-consistent simulations.  Finally, a test-particle realization
of the orbital distribution of the disk is generated
self-consistently according to the total galaxy potential for the models 
using the technique of Kuijken \& Dubinski (1995).
The test particles are then integrated in the time-dependent
gravitational field of the
two rigid potentials moving along the pre-calculated collision trajectory.
This technique produces tidal tail morphologies and kinematics remarkably
similar to self-consistent simulations, so we apply it generally to
models in our survey.

\begin{figure}
\plotone{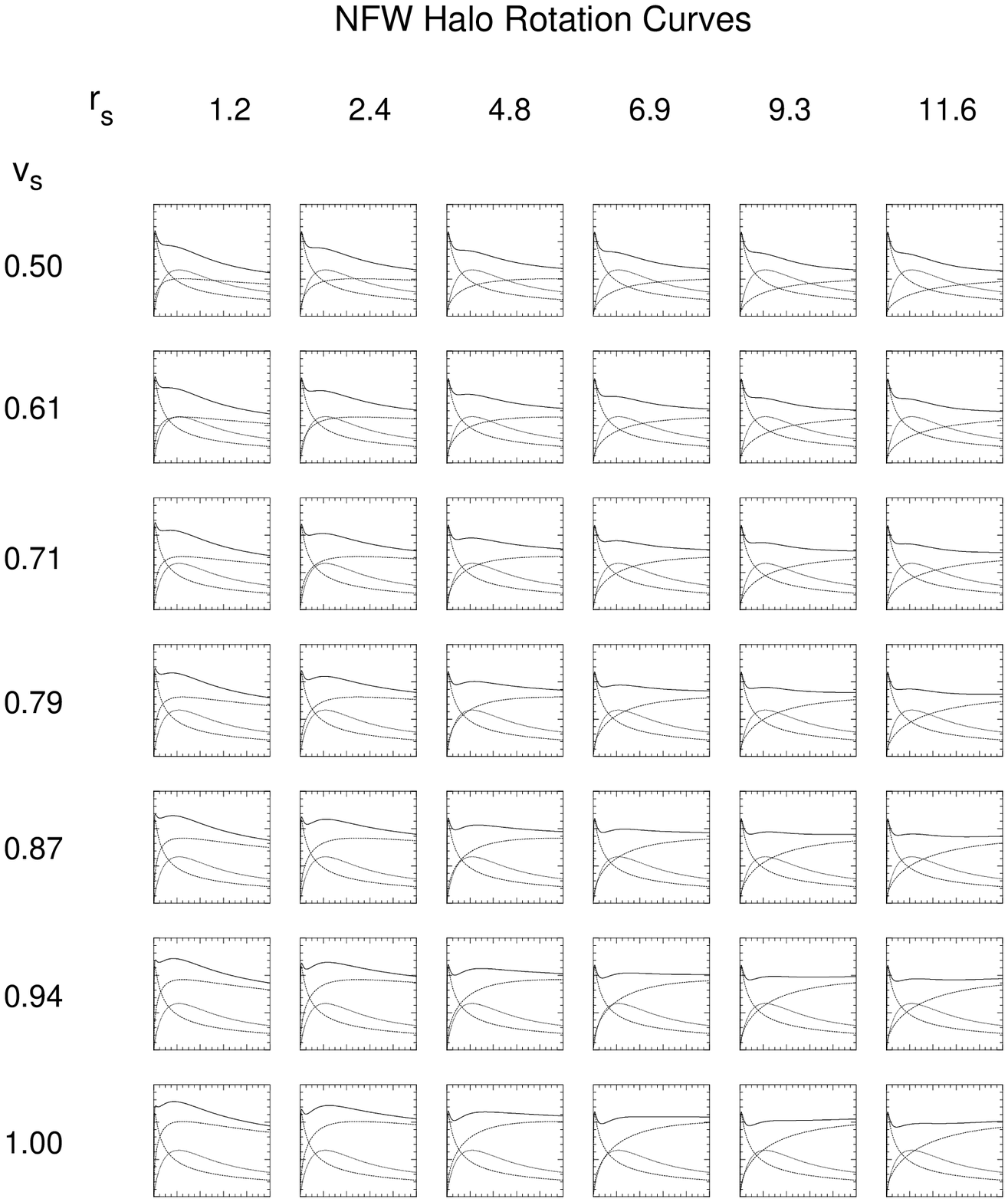}
\caption{Rotation curves of the galaxy models in the collision survey.
The contribution from the disk, bulge and halo as well as the net rotation
curve are shown within 10 scale lengths for each model.  A range of models
is covered with compact low-mass halos (upper left corner) to high-mass,
extended halos (lower right corner).}
\label{fig-2}
\end{figure}

Halos are parameterized in our study by
their scale length and circular velocity maximum.  We have chosen
six halo scale-lengths ranging from $r_s=1.2$ to 11.6
and seven circular velocities ranging from $v_s = 0.5$ to 1.0 for a total
of 42 models in the NFW study.  We also looked at a similar set of 42 H models.
The disk
exponential scale length and mass are both unity, so the disk's peak
circular velocity is $v_d = 0.62$ in these units.  The bulge has a mass of
$0.5 M_d$ and is the main contributor to the rotation curve within 1 scale
length.  This range of
models broadly covers the inferred properties of dark
halos around exponential galactic disks of Hubble type Sa-b
similar to the Milky Way.
Figure \ref{fig-2} shows the rotation curves within 10 scale-lengths.

\begin{figure}
\plotone{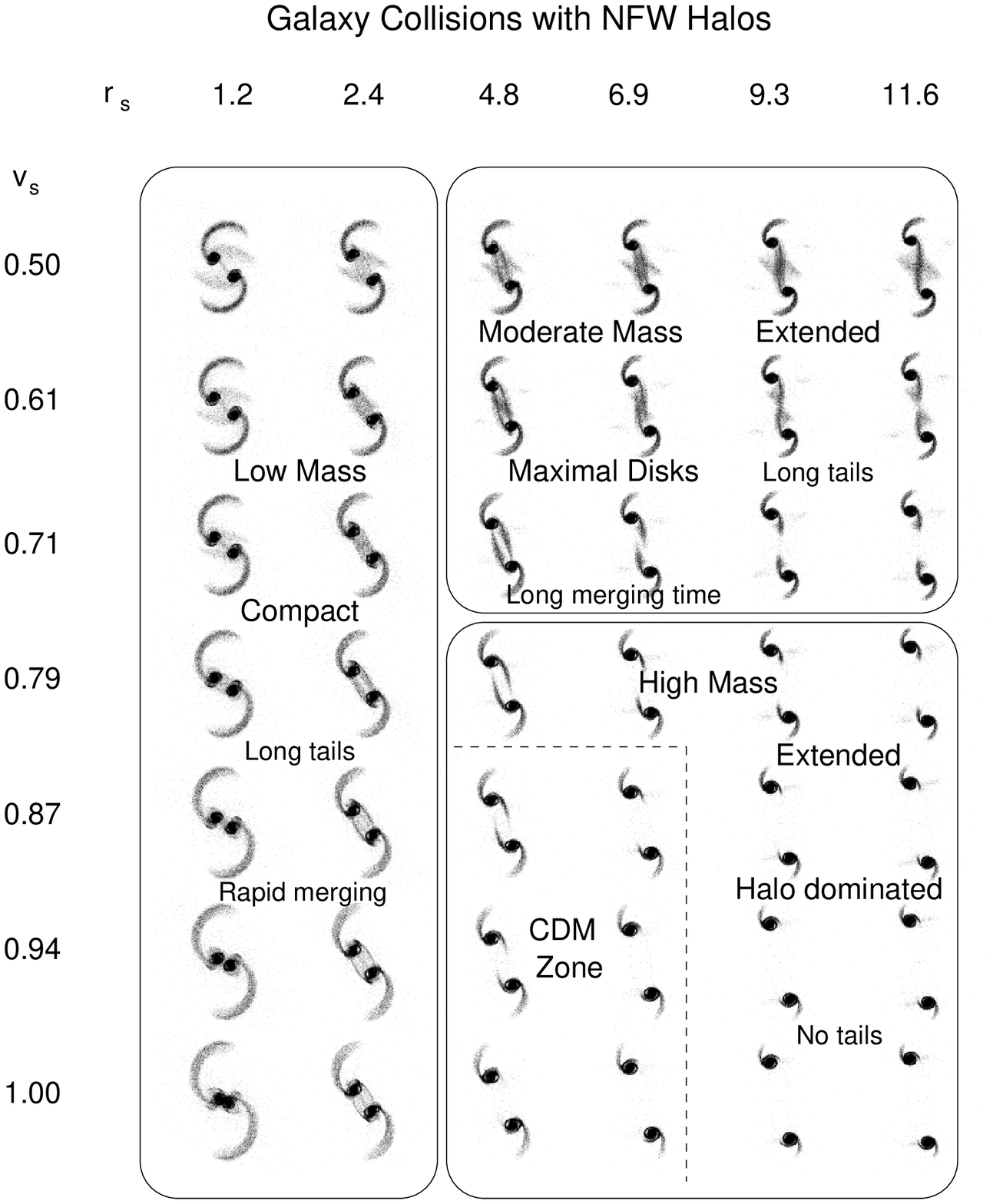}
\caption{Tidal tails resulting from galaxy collisions in the model survey.
See the text for a detailed description.}
\label{fig-3}
\end{figure}

Figure \ref{fig-3} presents the results of a series of planar, prograde
galaxy collisions.  
Each collision is shown at the same time, $t=30$, 
after pericentric
passage, corresponding to 5 disk orbital times at one scale length
or 500 million years for the Milky Way.  
The size of each box is 80 scale
lengths or about 300 kpc for the Milky Way.
The main results are that galaxies with 
low-mass, compact halos expel long tidal tails in collisions 
similar to those in observed interacting pairs 
while galaxies with high-mass extended halos 
expel very short tails unlike the Antennae or NGC 7525.
The region where halos are
similar to CDM predictions also produce relatively short tails in accord
with our earlier conclusions.  Galaxies with extended 
halos having smaller peak circular velocities (upper
right corner) can also produce long tidal tails and bridges since the
galaxies separate significantly after their encounter.  The potential is
shallow enough to allow long tails to fly off to large distances
but the low central density
of the halo does not provide enough dynamical friction to slow down the
galaxies significantly during their encounter in contrast to the
models in the lower left corner where the central halo density is much higher.  
The interacting pair
Arp 295 with its wide separation and long tails and connecting bridge
has a similar morphology to galaxies in the upper right corner.

\section{Discussion and Conclusions}

This more extensive survey
confirms our previous conclusion that collisions of galaxies with
dark halos like CDM halos are ineffective at making 
long tidal tails.  Halos which are compact and centrally concentrated 
with a range of circular velocities all make long tidal tails 
but extended halos can  only make long tails if they have a small
peak velocity i.e. a smaller contribution than the disk and bulge to the
inner rotation curve.
The successful galaxy models with 
extended halos are probably most similar 
to maximal and Bottema (1997) disk/halo models which are preferred by some
rotation curve analysts.
The main cosmological implication is that
CDM and critical density universe models are not favored because their dark
halos inhibit the formation of long tidal tails.  More compact
dark halos are predicted in low density universes, however, similar to models
in the lower left of Figure 2 (Navarro et al. 1997).

The disagreement with the relatively large mass estimates in the Milky Way
and external galaxies made through studies of satellite kinematics 
is harder to understand (e.g. Zaritsky et al. 1989; Kochanek 1996; Zaritsky
\& White 1994; Zaritsky et al. 1997).
Perhaps the difference is due to a selection effect -- galaxies chosen for
satellite kinematics studies are well-isolated while interacting galaxies
are obviously in pairs and not isolated.  The different halos may
reflect some cosmological variance resulting from differences in environment
or initial conditions.
In any case, the lower bounds of the confidence limits on halo masses 
from satellites
overlap with range of galaxy models that make long tidal tails.
Clearly, the systematic errors of these mass estimates should
be studied in more detail, perhaps through comparison to cosmological
simulations with adequate resolution to resolve satellites within their
galactic halos.   Galaxy interaction studies should also take their initial
conditions from cosmological models instead of the idealized models
examined
here.  Perhaps a consistent picture  will be found soon and the tails
will stop wagging the dogs.

%


%

\end{document}